\documentstyle[aps,prl,psfig,multicol,epsf,epsfig]{revtex}
\draft
\newcommand{\be}{\begin{equation}}
\newcommand{\ee}{\end{equation}}
\newcommand{\bea}{\begin{eqnarray}}
\newcommand{\eea}{\end{eqnarray}}

\def\ba{\begin{eqnarray}}
\def\ea{\end{eqnarray}}

\begin{document}
\title{Detect $\Delta G$ at BNL-RHIC via Double Quarkonium 
Production\\[2mm]} 
\author{Jiro Kodaira
\footnote{E-mail: kodaira@theo.phys.sci.hiroshima-u.ac.jp}
\, and \, Cong-Feng Qiao\footnote{
E-mail: qiao@theo.phys.sci.hiroshima-u.ac.jp}}
\address{Department of Physics, Hiroshima University\\
Higashi-Hiroshima 739-8526, Japan}
\maketitle

\begin{abstract}
The double spin asymmetry for exclusive $J/\psi$ pair 
production in the polarized $p-p$ collisions at RHIC is 
investigated. Our study shows that the asymmetry in this 
process is measurable at RHIC-SPIN experiments in the near 
future, and, hence it can be used to determine the gluon 
distribution function $\Delta G (x)$ in the polarized nucleon.
 

\end{abstract}

\begin{multicols}{2}

Since the first measurement of the polarized structure
function by the European Muon Collaboration \cite{emc}, 
an enormous amount of researches have been done on 
the nucleon spin structure both experimentally and 
theoretically. The unpolarized deep inelastic scattering 
(DIS) experiments show that the gluon shares a large 
portion of the parent proton's momentum. However, how 
the gluons share the spin of proton is still an open 
question. In the conventional DIS process, the gluon 
contributions are the Quantum Chromodynamics(QCD) 
higher order effects and we can not avoid some ambiguities 
coming from {\it e.g.} the factorization scheme dependence, 
when determining the polarized gluon distributions. 
Although there are some efforts \cite{GS,GRVO} to parameterize 
them, it seems difficult to obtain clear understanding of the 
role of gluons. 

Therefore, it is desirable to study other processes than DIS
in which a direct measurement of the polarized gluon is possible.
It is now expected that the polarized proton-proton collisions
(RHIC-Spin) at BNL relativistic heavy-ion collider RHIC will 
provide copious experimental data to unveil the polarized parton 
distributions. Since the major emphasis and strength of RHIC-Spin 
is to measure the gluon polarization, it is important and 
interesting to investigate various processes which are 
attainable experimentally to this aim.

Up to now, three processes are considered to be promising for 
measuring the polarized gluons, which are thought to be feasible 
at RHIC technically. Those are
\begin{itemize}
\item{High-$p_T$ Prompt Photon Production}
\item{Jet production}
\item{Heavy Flavor Production}
\end{itemize}
There are some advantages and disadvantages in each of these 
processes. The prompt photon production with polarized beams 
at RHIC is a useful process, but we must know precisely the 
quark distributions from the polarized DIS experiments.
For jet and heavy flavor production, the effects of 
hadronization and higher order corrections are not yet well 
controlled theoretically. For detailed discussions, see recent 
review paper in Ref. \cite{bssv}. Therefore, to develop more 
practical ways for measuring the gluon polarization at RHIC 
experiments is one of the urgent theoretical tasks today.

The quarkonium production and decays have long been taken as 
ideal processes to investigate the nature of QCD. Due to the 
approximately non-relativistic nature, the heavy quark and 
antiquark bound state is considered as one of the simplest 
system in QCD. The very clean signals of quarkonium leptonic 
decays lead to the experimental observations with a high 
precision, and therefore the quarkonium may play a crucial 
role in investigating many quantities and phenomena, {\it e.g.} 
the parton distribution, the QGP and even new physics. 

However, it should be pointed out that although the heavy 
quarkonium physics has been investigated for about thirty 
years, theoretical description of quarkonium production is 
still premature. Conventionally, the so-called color-singlet 
model (CSM) was widely employed in the study of heavy quarkonium
production and decays \cite{t.a.degrand}. In CSM, it is assumed 
that the $Q\bar{Q}$ pair produced in a high 
energy collision will bind to form a given quarkonium state
only if the $Q\bar{Q}$ pair is created in color-singlet 
state with the same quantum numbers as the produced bound 
states; as well, in the quarkonium decays the annihilating 
$Q\bar{Q}$ pair will be in short distance and singlet with 
the same quantum numbers as its parent bound states.
It is assumed in CSM that the production amplitudes 
can be factorized into short distance and long distance 
parts. The short distance sector is perturbative QCD
applicable, while all the long distance nonperturbative 
effects are attributed to a single parameter, the wave 
function. Nevertheless, color-singlet factorization is only 
an ad hoc hypothesis. There are no general arguments to 
guarantee such a naive assumption to be held up to higher 
order radiative corrections. Recently, from the Fermilab 
Tevatron data collected in the 1992-1993 run, CDF group 
found \cite{cdf1,cdf2,cdf3} that the large-$p_T$ $J/\psi(\psi')$ 
were produced much more than the leading order(LO)   
CSM predictions. This phenomenon was referred as  
"$\psi$-surplus" production or "-anomaly".

In order to explain the experimental results of $\psi$ surplus 
production, the color-octet mechanism(COM)\cite{fleming} was 
proposed based on a novel effective theory, the non-relativistic 
QCD(NRQCD) \cite{nrqcd}. According to COM the produced $c\bar{c}$ 
pair in color-octet configuration may also evolve into $\psi$ 
via radiation or absorption of soft gluons. For details, 
readers are recommended to reference \cite{nrqcd}. The 
appearance of this new quarkonium production mechanism 
brought deep effects on the quarkonium physics. Having 
achieved the first-step success in explaining the 
CDF data, the color-octet mechanism encounters 
also difficulties in other processes \cite{rothstein}. 
As discussed in Ref. \cite{qcf1}, the extent of importance 
of color-octet contributions in charmonium production remains 
unfixed for the time being. Therefore, to use quarkonium as probes
to investigate new physics, one should really take the large 
ambiguities remaining in color octet matrix elements and 
higher order corrections into consideration, especilly
for charmonium system.

In literature, a series of efforts have been made so far for 
measuring the polarized parton distributions through quarkonium 
production processes \cite{bs,cp,dr,mty,jk}. Unfortunately, most 
of previous investigations were not directly applied to the RHIC 
experiments and are spoiled by the uncertainties aforementioned. 
In this work, we show that the exclusive double heavy quarkonium 
production in the polarized proton-proton collision would be one 
of the ideal processes to measure the polarized gluon distributions
and may play at least a supplemental role to the presently proposed 
program at RHIC. Due to being an exclusive process that can not be 
realized in COM, the uncertainties remaining in color-octet matrix
elements are ruled out for our concern, which enables us to make 
more precise predictions. In addition, the double quarkonium 
production also has several advantages in reducing theoretical 
uncertainties from other sources. That is, (1) The concerned 
quantity, the asymmetry, is almost independent of the relativistic 
corrections and non-perturbative uncertainties, which is normally 
not negligible in the study of charmonium system. The reason is 
that these kinds of uncertainties appearing in both numerator 
and denominator in the definition of asymmetry may cancel 
out. (2) The higher order QCD corrections can be well controlled by 
applying a suitable $p_T$ cut for the charmonium system, because the
the strength of strong interaction here highly correlates to the 
transverse momenta of final states. (3) At RHIC energy, especially 
for $p-p$ collision, the $q \bar{q}$ annihilation process is 
reasonably negligible compared to the prevailing gluon-gluon 
fusion process. Therefore, the double-quarkonium stands as a 
very sensitive system for measuring gluon polarizations.

Years ago, there were discussions of observing the gluon polarization 
by means of exclusive double $J/\psi$ production \cite{bj,gm}. 
However, the main concerns there were not on RHIC physics, and 
analytical expressions were not presented for more extensive use. 

\vskip 2mm
\begin{figure}
\begin{center}
\psfig{file=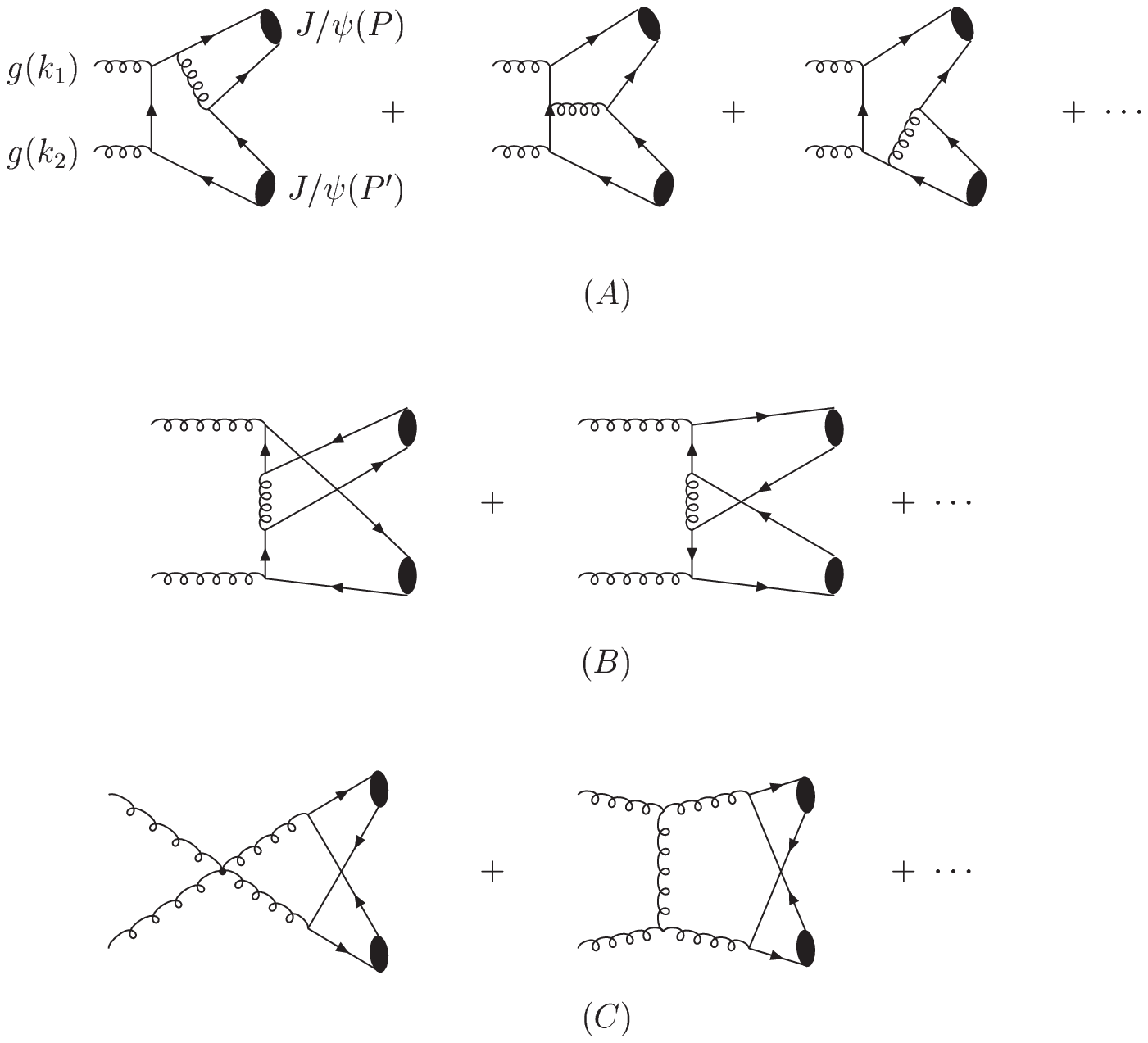, bbllx=70pt,bblly=280pt,bburx=500pt,
bbury=672pt,width=8cm,height=6cm,clip=0}
\end{center}
\caption[]{Typical Feynman diagrams for $g\, + \,g \rightarrow
J/\psi + J/\psi$. The topological groups (A) and (B) show up 
in the photon-photon to double charmonium production process. 
The group (C) is necessary in guaranteeing the gauge invariance.}
\end{figure}

At the partonic level, the concerned process is similar to 
the photon-photon scattering in double quarkonium production 
\cite{qcf3}. Because of the non-abelian nature of QCD, 
here we should consider more diagrams as show in Figure 1. 
When calculating the helicity-dependent matrix elements, one
must project onto definite helicity states for the initial gluons.
In our calculation, this is achieved by taking the ghost free 
expression for the gluon state with helicity $\lambda = \pm$ ,
\bea
\epsilon_{\mu}(k_1,\,\lambda)\;\epsilon^\star_{\nu}(k_1,\,\lambda)
&=&\frac{1}{2}\left[-g_{\mu\nu} + \frac{k_{1\mu} k_{2\nu} + k_{1\nu} 
k_{2\mu}} {k_1 \cdot k_2}\right. \nonumber \\
&+& \left. i \,\lambda\;
\epsilon_{\mu\nu\alpha\beta}\frac{k_{1\alpha} k_{2\beta}}
{k_1 \cdot k_2}\right]
\; .
\label{eq0}
\eea

The double spin asymmetry $A$ for $J/\psi$ pair production is 
defined as,
\bea
A &=& \frac{d \sigma (p_+ p_+ 
\rightarrow J/\psi J/\psi) - d \sigma (p_+ p_- 
\rightarrow J/\psi J/\psi)}{d \sigma (p_+ p_+ 
\rightarrow J/\psi J/\psi) + d \sigma (p_+ p_- 
\rightarrow J/\psi J/\psi)}\nonumber \\
&=& \frac{E d \Delta\sigma/d^3p}{E d \sigma/d^3p}\; ,
\label{eq1}
\eea
where $p_+$ and $p_-$ denote the helicity states of the incident 
protons being positive and negative, respectively. Here and in the 
following, the parity conservation is taken for granted. In terms 
of the gluon densities and the partonic cross sections, this 
asymmetry reads,
\be A 
     =  \frac{ \int dx_1 dx_2  d \Delta \hat{\sigma}
             \Delta G(x_1, Q^2 ) \Delta G(x_2, Q^2 )}
         {\int dx_1 dx_2  d\hat{\sigma} G(x_1, Q^2 ) G(x_2, Q^2)} \; ,
\ee
where $\Delta G(x, Q^2) = G_+ (x, Q^2 ) - G_- (x, Q^2)$ and 
$G(x, Q^2) = G_+ (x, Q^2) + G_- (x, Q^2)$
are the polarized and unpolarized gluon distributions defined at the
scale $Q^2$. The unpolarized (polarized) partonic cross section
$\hat{\sigma}$ ($\Delta \hat{\sigma}$) is defined as,
\be 
\hat{\sigma} = \frac{1}{4}\ \sum_{\lambda,\lambda'}\ \hat{\sigma}\
(\lambda,\;\lambda'),\;
\Delta\hat{\sigma} = \frac{1}{4}\ \sum_{\lambda,\lambda'}\ 
\lambda\lambda' \ \hat{\sigma} \ (\lambda,\;\lambda').
\ee
The concerned process, as schematically presented in Figure 1, 
gives the polarized partonic differential cross section as
\bea
&&\frac{d\Delta \hat{\sigma}}{dt}=\frac{-16 \alpha_s^4 \pi |R(0)|^4}
{81 s^8 (m^2 - t)^4 (m^2 - u)^4}
\times \Bigl[ \ 2744 m^{24}\nonumber\\  
&-& 15240 m^{22} (t + u)\nonumber\\
&+& m^{20} (32110 t^2 + 90076 t u + 32110 u^2) \nonumber\\ 
&-& 16 m^{18} (2025 t^3 + 12673 t^2 u + 12673 t u^2 + 2025 u^3) 
\nonumber\\ 
&+& 2 t^4 u^4 (349 t^4 - 908 t^3 u + 1374 t^2 u^2 - 908 t u^3 + 349 u^4) 
\nonumber\\
&+& 4 m^{16} (3903 t^4 + 57292 t^3 u + 117766 t^2 u^2 + 57292 t u^3 
\nonumber\\ 
&+& 3903 u^4) - 4 m^{14} (510 t^5 + 36713 t^4 u + 135685 t^3 u^2 
\nonumber\\
&+& 135685 t^2 u^3 + 36713 t u^4 + 510 u^5)\nonumber\\
&+& m^{12} (-1461 t^6 + 58600 t^5 u + 364313 t^4 u^2 \nonumber\\
&+& 594840 t^3 u^3 + 364313 t^2 u^4 + 58600 t u^5 - 1461 u^6) 
\nonumber\\
&+& 4 m^2 t^2 u^2 (9 t^7 - 505 t^6 u + 44 t^5 u^2 - 556 t^4 u^3
\nonumber\\ 
&-& 556 t^3 u^4 + 44 t^2 u^5 - 505 t u^6 + 9 u^7) \nonumber\\
&+& 2 m^{10} (381 t^7 - 7111 t^6 u - 83783 t^5 u^2 - 180639 t^4 u^3
\nonumber\\ 
&-& 180639 t^3 u^4 - 83783 t^2 u^5 - 7111 t u^6 + 381 u^7)
\label{eq2} \\
&+& m^8 (-79 t^8 + 1272 t^7 u + 54526 t^6 u^2 + 156224 t^5 u^3 
\nonumber\\ 
&+& 163850 t^4 u^4 + 156224 t^3 u^5 \nonumber\\
&+& 54526 t^2 u^6 + 1272 t u^7  - 79 u^8) \nonumber\\
&+& m^4 t u (-36 t^8 + 1471 t^7 u + 9764 t^6 u^2 + 12863 t^5 u^3 
\nonumber\\ 
&+& 7196 t^4 u^4 + 12863 t^3 u^5 + 9764 t^2 u^6 \nonumber\\
&+& 1471 t u^7 - 36 u^8) \nonumber\\
&-& 2 m^6 (2 t^9 + 17 t^8 u + 5151 t^7 u^2 + 25947 t^6 u^3\nonumber\\
&+& 24439 t^5 u^4 + 24439 t^4 u^5 + 25947 t^3 u^6\nonumber\\
&+& 5151 t^2 u^7 + 17 t u^8 + 2 u^9) \ \Bigr] \; ,\nonumber
\eea
where $R(0)$ is the wave function of $J/\psi$ at the origin
and $m$ is the rest mass of $J/\psi$;\  $s,\,t$ and $u$ are the 
Mandelstam variables for the partonic system. The sum of our 
results for different helicity combinations gives the unpolarized 
partonic differential cross section, $\hat{\sigma}$ in (4), which 
agrees with the results in Refs. \cite{qcf2,hm}.

In the following, we investigate the spin asymmetry in the RHIC-Spin 
experiment at $\sqrt{s} = 500$ GeV which is the planed highest RHIC 
energy. In our numerical calculations, the scale $Q^2$ of the parton 
distribution function and the strong coupling constant is taken
to be the transverse momentum of $J/\psi$ for the $p_T$ distributions.
Whereas, in the calculation of the angular distribution of the spin 
asymmetries and the integrated cross sections, the scale is taken to 
be $Q^2 = m^2$. The nonrelativistic relation $m = 2\ m_c$, with
$m_c = 1.5$ GeV, is used and $|R(0)|^2\ = \ 0.8\ \rm{GeV}^3$. 

\begin{figure}[tbh]
\begin{center}
\epsfig{file=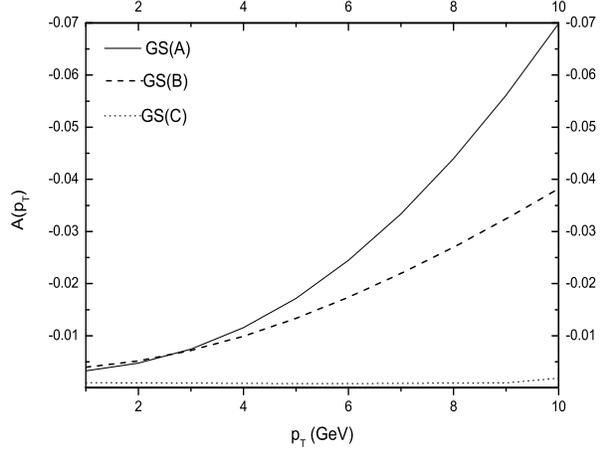,bbllx=30pt,bblly=710pt,bburx=330pt, 
bbury=912pt,width=8cm,height=6cm,clip=0}
\end{center}
\caption[]{Spin asymmetry of $J/\psi$ pair production in color-singlet
model versus transverse momentum for different set of GS polarized 
parton distributions at RHIC with colliding energy $\sqrt{s} = 500$ GeV.}
\label{graph1}
\end{figure}

We plot in Figure 2 the double spin asymmetry versus the transverse 
momentum of $J/\psi$ with respect to the proton beams with different 
sets of the GS parameterizations \cite{GS}, and in Figure 3 the angular 
distribution of the asymmetry in the parton center-of-mass frame. 
The charmonium pair exclusive production happens only through 
the color-singlet scheme as shown in Figure 1 and so it is feasible 
to reconstruct the parton center-of-mass system in experiment. 
To be consistent with the use of the GS polarized distributions, 
we use the MRST parameterization \cite{mrst} for the unpolarized 
gluon distribution. From results shown in Figures 2 and 3, 
it is clear that the asymmetries obtained with different 
parameterizations diverse quite much. 

We notice that to observe the spin asymmetry both in 
$p_T$ and angular distributions, a high luminosity is required.
Nevertheless, having a relatively high accuracy in theory,
even with a few experimental events our results may tell 
something about the gluon polarization inside the proton. 
From diagrams 2 and 3 it is  easy to figure out that the 
charmonium pairs are produced mainly in low $p_T$ and 
the forward direction relative to the beams, where the 
set C gives a small asymmetry compared to sets A and B. 

\begin{figure}[tbh]
\begin{center}
\epsfig{file=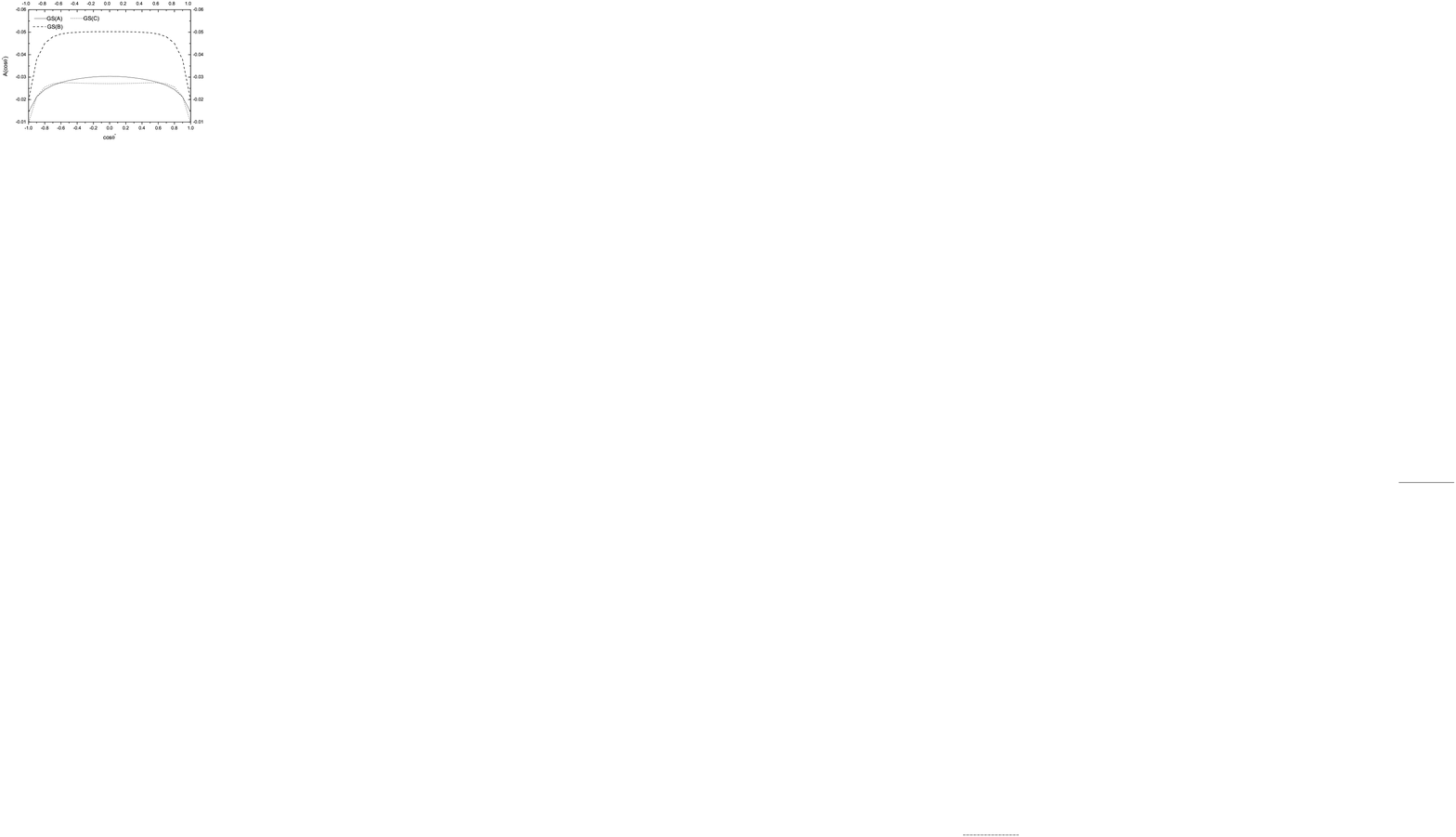,bbllx=20pt,bblly=1024pt,bburx=325pt,
bbury=1230pt,width=8cm,height=6cm,clip=0}
\end{center}
\caption[]{Angular differential asymmetry distribution of the 
$J/\psi$ pair in the parton center-of-mass frame at RHIC with
colliding energy $\sqrt{s}$ = 500 GeV.}
\label{graph2}
\end{figure}    

In order to estimate the event rate for the $J/\psi$ pair productions, 
we calculate the total cross section with different parton
parameterizations. The results are collected in table I. As expected, 
the discrepancies among these predictions are not very large. 
Here, the notation $\sigma_{\mu^+\mu^-}$ means that the branching 
ratio of $B(\psi \rightarrow {\mu^+\mu^-}) =  0.0588$, as the 
practical measuring mode to reconstruct the charmonium state, 
is included. From the predicted cross sections, we see
that with the integrated luminosity of $800\;\rm{pb}^{-1}$ 
in the future run of RHIC, there will be thousands of $J/\psi$
pair events to be detected, which can certainly give
us some information on the gluon polarization in the nucleon.

\vskip 6mm
{\small TABLE I. Total cross sections for $J/\psi$ pair production
at RHIC with $\sqrt{s} = 500$ GeV, evaluated with different parton 
distributions.}

\begin{center}
\begin{tabular}{ccc}\hline\hline            
\rule[-1.2ex]{0mm}{4ex}&\hspace{-0.6cm} 
$\sigma^{tot}_{\mu^+\mu^-}$ \hspace{0.6cm}&
$\sigma_{\mu^+\mu^-}(|p_T|>1 \;\mbox{GeV})$ \\ \hline
\rule[-1.2ex]{0mm}{4ex} CTEQ5L\cite{cteq} 
\hspace{0.6cm} & $11.8 \; \mbox{pb}$ \hspace{0.6cm}
& $7.3\; \mbox{pb}$ \\ \hline
\rule[-1.2ex]{0mm}{4ex} MRST \cite{mrst}\hspace{0.6cm} & $6.5 
\; \mbox{pb}$\hspace{0.6cm} 
& $4.3\; \mbox{pb}$ \\ \hline
\rule[-1.2ex]{0mm}{4ex} GRV \cite{grv}
\hspace{0.6cm} & $7.4 \; \mbox{pb}$ \hspace{0.6cm}
& $4.7\; \mbox{pb}$ \\ \hline\hline
\end{tabular}
\end{center}
\vskip 5mm

To conclude, in this work we have shown that the exclusive $J/\psi$ 
(quarkonium) pair production at RHIC may stand as a novel process in 
measuring the gluon spin distributions inside the polarized nucleon. 
The large mass of heavy quark guarantees that perturbative calculation 
is applicable to this process; the asymmetry, rather than cross 
sections, eliminates large amount of uncertainties which come from the 
non-perturbative hadronization and relativistic corrections. 
The higher order QCD corrections may be properly controlled by 
employing a suitable $p_T$ cut. We have also discussed the 
feasibility of observing the double spin asymmetry via this
process. Our results show that for the time being the accumulated 
data with low achievable fraction of polarization in the RHIC 
beam are not yet enough to analyze this process for the purpose 
of measuring the gluon helicity distributions. However, in the 
future run with colliding energy of 500 GeV and the accumulated 
luminosity 800 $\rm{pb}^{-1}$, the $J/\psi$ pair events can be 
surely detected and the gluon polarization could be measured. 
With the expected upgrade of RHIC in future, the $J/\psi$ pair 
production may become a promising process with very clean signal 
and less theoretical uncertainties in uncovering the nucleon 
spin structures. 

Finally, it should be mentioned that there are two kinds of 
backgrounds which may interfere with the measurement of our 
proposed process. The first is the contribution of higher 
excited quarkonium states feeddown to double $J/\psi$. Which 
is not negligibly small generally speaking, however, for 
pair production they are doubly suppressed. The second is 
the quakonium pairs production through color-octet mechanism, 
which is known to be the dominant $J/\psi$ pair production 
scheme at high energy and large $p_T$ via $g g \rightarrow 
g g$ hard interaction and with both final state gluons 
fragmenting to color-octet intermediate states and then 
evolving into qaurkonia nonperturbatively. Nevertheless,
in both cases the $J/\psi$ pairs are not exclusively produced, 
hence, experimentally they can be excluded with care.

\vskip 5mm
The work of J.K. was supported in part by the 
Monbu-kagaku-sho Grant-in-Aid for Scientific Research 
No.C-13640289. The work of C-F.Q. was supported by the 
Grant-in-Aid of JSPS committee. Authors would like to thank
M.G. Perdekamp and K. Hagiwara for helpful comments.

\end{multicols}
\end{document}